\documentclass{rspublic}
\usepackage{amssymb,amsmath}

\begin{document}

\title[Integrability and Linearization]{On the complete integrability and
linearization of nonlinear
ordinary differential equations - Part IV: Coupled second order equations}

\author[Chandrasekar, Senthilvelan and Lakshmanan]{V. K. Chandrasekar,
M. Senthilvelan and M. Lakshmanan}

\affiliation{Centre for Nonlinear Dynamics, Department of Physics,
Bharathidasan Univeristy, Tiruchirapalli - 620 024, India}

\label{firstpage}

\maketitle

\begin{abstract}{Nonlinear differential equations, Coupled second order, Integrability, Integrating factors, Uncoupling}

Coupled second order nonlinear differential equations are of fundamental importance in dynamics. In this part of our study on the integrability and linearization  of nonlinear ordinary differential equations we focus our attention on the method of deriving general solution of two coupled second order nonlinear ordinary differential equations through the extended Prelle-Singer procedure.  We describe a procedure to obtain integrating factors and required number of integrals of motion so that the general solution follows straightforwardly from these integrals.  Our method tackles both isotropic and non-isotropic cases in a systematic way.  In addition to the above, we introduce a new method of transforming coupled second order nonlinear ODEs into uncoupled ones.  We illustrate the theory with potentially important examples.

\end{abstract}

\section{Introduction}
In this part of our study on the integrability and linearization of
nonlinear ordinary differential equations (ODEs) we focus our
attention on the theoretical formulation and applications of the
modified Prelle-Singer (PS) procedure (Prelle \& Singer 1983; Duarte 
\textit{et al.} 2001;
Chandrasekar \textit{et al.} 2005$a$; 2006) to a set of two 
coupled second
order ODEs. The need for this demonstration is due to the fact that
classifying and studying two degrees of freedom dynamical systems are 
highly nontrivial problems in the theory of nonlinear dynamical
systems. Historically, several techniques have been proposed to
identify and obtain general solutions of two coupled second
order ODEs.  To cite a few we mention Painlev\'e analysis, Lie
symmetry analysis, generalized Noether symmetries technique, direct
method and so on (Ramani \textit{et al.} 1989; Lakshmanan \&
Sahadevan 1993; Bluman \& Anco 2002; Lakshmanan \& Rajasekar 2003). Each of these methods have their
own advantages and limitations.
For example, among the above, certain methods fulfil necessary
conditions alone whereas the others guarantee only sufficient
conditions for the complete integrability of the system concerned.
This factor alone is a motivating factor to search for 
more and more powerful methods to isolate and classify integrable
and non-integrable dynamical systems. In this direction, by
extending the PS procedure and its applications to coupled second
order ODEs we argue that the PS method can be used as a stand-alone
technique to solve a wide class of ODEs of any order irrespective of
whether it is a single or coupled equation.

We mention here that the present analysis is not a straightforward
extension of the scalar case.  In fact by prolonging the theoretical
formulation to the coupled second order ODEs, we deduce the
determining equations for the integrating factors and null forms
appropriately such that one can obtain the aforementioned functions
in a more efficient and straightforward way.  Thus the method of
obtaining the integrating factors for the given equation is also
augmented in this procedure in an efficient manner. Further, while
studying the coupled dynamical systems one may face both isotropic
and non-isotropic cases.  Our method covers both of them in a
natural way. In addition to the above, in this paper, we also
introduce a new method to transform two coupled second
order ODEs to two uncoupled second order ODEs. 
Thus, the PS procedure inherits several remarkable features both at the theoretical foundations as well as in the range of applications which we have listed out already in Chandrasekar \textit{et al.} (2005$a$). Finally, we note
that we have carefully fixed the examples so that the basic features
associated with this method and the results which it leads to could be
explained in an efficient way.

The plan of the paper is as follows. In \S2 we describe the PS
method applicable for coupled second-order ODEs and indicate the new
features in finding the integrating factors and integrals of motion.
In \S3 we establish a connection between integrating factors and the
form of equations. In \S4, the uncoupled equations are briefly considered. 
In \S5,  we discuss elaborately the method of constructing integrals and
general solutions for the coupled nonlinear ODEs. We support the theory with two nontrivial examples which are being discussed in the contemporary literature. 
We also briefly discuss the
application of our procedure for the case of Liouville integrable
systems in \S5d.  We devote \S6 to demonstrate yet another method to identify transformation variables from the first integrals which can be effectively used to rewrite the system of coupled ODEs into uncoupled ones so that one can integrate the resultant equation easily and obtain the general solution. We present our
conclusions in \S7.

\section{Prelle-Singer method for coupled second order ODEs}
\label{sec2}
\subsection{General Theory}
Let us consider a system of two coupled second order ODEs of the form
\begin{eqnarray}
\ddot{x}=\frac{d^2x}{dt^2}=\frac{P_1}{Q_1}, \quad
\ddot{y}={\frac{d^2y}{dt^2}}={\frac{P_2}{Q_2}}, \quad { P_i,Q_i}\in
C{[t,x,y,\dot{x},\dot{y}]},\; i=1,2,\label {cso01}
\end{eqnarray}
where $P_i$ and $Q_i$ are analytic functions of the variables $t,x,y,\dot{x}$
and $\dot{y}$. Let us suppose that the system (\ref{cso01}) admits a
first integral of the form $I(t,x,y,\dot{x},\dot{y})=C$
 with $C$ constant on the solutions so that the total differential gives
\begin{eqnarray}
dI={I_t}{dt}+{I_{x}}{dx}+{I_{y}}{dy}+{I_{\dot{x}}}{d\dot{x}}
+{I_{\dot{y}}}{d\dot{y}}=0, \label {cso02}
\end{eqnarray}
where subscript denotes partial differentiation with respect to that variable.
Rewriting ({\ref{cso01}) in the form
\begin{eqnarray}
\frac{P_1}{Q_1}dt-d\dot{x}=0,\qquad
\frac{P_2}{Q_2}dt-d\dot{y}=0 \label {cso03}
\end{eqnarray}
and adding null terms
$s_1(t,x,y,\dot{x},\dot{y})\dot{x}dt
- s_1(t,x,y,\dot{x},\dot{y})dx$ and $s_2(t,x,y,\dot{x},\dot{y})\dot{y}dt
- s_2(t,x,y,\dot{x},\dot{y})dy$
 with the first equation in (\ref{cso03}), and
$u_1(t,x,y,\dot{x},\dot{y})\dot{x}dt  -
u_1(t,x,y,\dot{x},$ $\dot{y})$ $dx $ and $u_2(t,x,y,\dot{x},\dot{y})\dot{y}dt -
u_2(t,x,y,\dot{x},\dot{y})dy $ with the second equation in (\ref{cso03}),
respectively, we obtain that, on the solutions, the 1-forms
\begin{subequations}
\begin{eqnarray}
&&(\frac{P_1}{Q_1}+s_1\dot{x}+s_2\dot{y})dt-s_1dx-s_2dy-d\dot{x}=0,\label {cso04}\\
&&(\frac{P_2}{Q_2}+u_1\dot{x}+u_2\dot{y})dt-u_1dx-u_2dy-d\dot{y}=0.\label {cso05}
\end{eqnarray}
\label {cso06}
\end{subequations}

At this stage, we wish to point out that one can also analyse the coupled second order ODEs (\ref{cso01}) by rewriting them as a set of four coupled first order ODEs of the form $\dot{x}=x_1,\;\dot{x}_1=P_1/Q_1,\;\dot{y}=y_1$, $\dot{y}_1=P_2/Q_2$ and its equivalent one forms. By introducing four integrating factors, one can deduce the relevant determining equations by following the procedure given by us in the earlier paper, Part-III (Chandrasekar \textit{et al.} 2008), to the above system of first order ODEs. However, after examining several examples we find that it is more advantageous to solve the system (\ref{cso01}) in the second order form itself rather then introducing more number of variables. The procedure is as follows.


Now, on the solutions, the 1-forms (\ref{cso02}) and
(\ref{cso06}) must be proportional. Multiplying (\ref{cso04}) by the factor
$ R(t,x,y,\dot{x},\dot{y})$ and (\ref{cso05}) by
the factor $ K(t,x,y,\dot{x},\dot{y})$, which act as the integrating
factors for (\ref{cso04}) and (\ref{cso05}), respectively, we have on the
solutions that
\begin{eqnarray}
dI=R(\phi_1+S\dot{x})dt+K(\phi_2+U\dot{y})dt-RSdx
-KUdy-Rd\dot{x}-Kd\dot{y}=0,\;\;\label {cso07}
\end{eqnarray}
where $ \phi_i\equiv {P_i}/{Q_i},\; i=1,2,$ and
$S=\frac{Rs_1+Ku_1}{R}$ and $U=\frac{Rs_2+Ku_2}{K}$. Comparing
equations (\ref{cso07}) and (\ref{cso02}) we have, on the solutions,
the relations
\begin{eqnarray}
 I_t  =R(\phi_1+S\dot{x})+K(\phi_2+U\dot{y}),\; I_{x}  = -RS,
  I_{y} =-KU,\;
 I_{\dot{x}}  =-R,\;
  I_{\dot{y}} =-K.
 \label {cso08}
\end{eqnarray}
The compatibility conditions between the equations (\ref{cso08}), namely $I_{tx}=I_{xt},\;I_{t\dot{x}}=I_{\dot{x}t},\;I_{ty}=I_{yt},\;I_{t\dot{y}}=I_{\dot{y}t},\;I_{xy}=I_{yx},\;I_{x\dot{x}}=I_{\dot{x}x},\;I_{y\dot{y}}=I_{\dot{y}y},\;I_{x\dot{y}}=I_{\dot{y}x},\;I_{y\dot{x}}=I_{\dot{x}y}$ and $I_{\dot{x}\dot{y}}=I_{\dot{y}\dot{x}}$, provide us
the conditions,
\begin{eqnarray}
D[S] &=&-\phi_{1x}-\frac{K}{R} \phi_{2x}
    +\frac{K}{R}S\phi_{2\dot{x}}
           +S\phi_{1\dot{x}}+S^2,  \label {eq23}\\
D[U] &=&-\phi_{2y}-\frac{R}{K}\phi_{1y}
    +\frac{R}{K}U\phi_{1\dot{y}}
           +U\phi_{2\dot{y}}+U^2,  \label {eq24}\\
D[R]  &=&-(R\phi_{1\dot{x}}+K\phi_{2\dot{x}}+R S),  \label {eq25}\\
D[K]  &=&-(K\phi_{2\dot{y}}+R\phi_{1\dot{y}}+K U), \label {eq26}\\
SR_{y} &=&-RS_{y}+UK_{x}+KU_{x},\;\;R_{x} =SR_{\dot{x}}+RS_{\dot{x}},
  \label {eq31}\\
R_{y} &=&UK_{\dot{x}}+KU_{\dot{x}},\qquad\qquad \;
K_{x} =SR_{\dot{y}}+RS_{\dot{y}},  \label {eq29}\\
K_{y}&=&UK_{\dot{y}}+KU_{\dot{y}},\qquad\qquad\;
R_{\dot{y}} =K_{\dot{x}}.\label {eq32}
\end{eqnarray}
Here the total differential operator, $D$, is defined by
$
D =\frac{\partial}{\partial{t}}+\dot{x}\frac{\partial}{\partial{x}}
+\dot{y}\frac{\partial}{\partial{y}}
+\phi_1\frac{\partial}{\partial{\dot{x}}}+\phi_2\frac{\partial}
{\partial{\dot{y}}}$. Integrating equations~(\ref{cso08}), we obtain the
integral of motion,
\begin{eqnarray}
I=r_1+r_2+r_3+r_4
-\int\bigg[K+\frac{d}{d\dot{y}}\bigg(r_1+r_2+r_3+r_4\bigg)\bigg]
d\dot{y},
\label {cso09}
\end{eqnarray}
where
\begin{eqnarray}
r_1&=\int\bigg(R(\phi_1+S\dot{x})+K(\phi_2+U\dot{y})\bigg)dt,
\qquad
r_2=-\int\bigg(RS+\frac{d}{dx}(r_1)\bigg)dx,&\qquad\nonumber\\
r_3&=-\int\bigg(KU+\frac{d}{dy}(r_1+r_2)\bigg)dy,\qquad
r_4=-\int\bigg[R+\frac{d}{d\dot{x}}\bigg(r_1+r_2+r_3\bigg)\bigg]d\dot{x}.&
\nonumber
\end{eqnarray}
Solving the determining equations,
(\ref{eq23})-(\ref{eq32}), consistently we can obtain expressions for the
functions $(S,U,R,K)$.
Substituting them into (\ref{cso09}) and evaluating the integrals we can constract
the associated integrals of motion.
It is also clear that equation~(\ref{cso01}) can be
considered as a completely integrable system once we obtain
four independent integrals of motion through this procedure.

\section{Connection between integrating factors and nature of equations}
We note that equations (\ref{eq23})-(\ref{eq32}) constitute an 
overdetermined system for the four unknowns, namely $S,U,R$ and $K$.
Among these equations, the first four equations,
(\ref{eq23})-(\ref{eq26}), constitute the evolution equations for
the variables $S,U,R$ and $K$. We mention here that combining
equations (\ref{eq23})-(\ref{eq26}) one can get the following two
identities, namely
\begin{eqnarray}
D{[R S]} =-(R\phi_{1x}+K\phi_{2x}),\quad D{[KU]} =-(R\phi_{1y}+K\phi_{2y}).
\label {eq33}
\end{eqnarray}
Any solution $(S,U,R,K)$ which satisfies the equations
(\ref{eq23})-(\ref{eq26}) also satisfies the equation (\ref{eq33}).
Once the functions $S,\;U,\;R$ and $K$ are fixed then the rest of
the problem is to verify whether these functions satisfy the `extra
determining equations', that is (\ref{eq31})-(\ref{eq32}), or not.
If these functions satisfy the extra determining equations then they
form a compatible set of solution and one can proceed to construct the
associated integral of motion from (\ref{cso09}). On the other hand 
if the functions do
not satisfy the extra determining equations  then one has to look
for alternate ways to obtain compatible solutions. In fact, in
practice, one often meets the case in which  certain solution(s)
which satisfies(satisfy) the evolutionary determining equations
(\ref{eq23})-(\ref{eq26}) does(do) not satisfy the extra determining
equations.   More specifically, for a class of problems one often
gets one or two or even three  sets of $(S,U,R,K)$ by solving
(\ref{eq23})-(\ref{eq26}) which does(do) satisfy the rest and
another(other) set(s) does(do) not satisfy the later equations. In
this situation we find an interesting fact that one can use the
integral(s) derived from the  set(s) which satisfies(satisfy) all
the equations (\ref{eq23})-(\ref{eq32})  and deduce the other
compatible solution(s) $(S,U,\hat{R},\hat{K})$  (definition of
$\hat{R}$ and $\hat{K}$ follows). For example, let the set
$(S_4,U_4,R_4,K_4)$ be a  solution of the evolution
equations~(\ref{eq23})-(\ref{eq26}) and not of the  extra
determining equations~(\ref{eq31})-(\ref{eq32}). After analyzing
several examples we find that one can make the set
$(S_4,U_4,R_4,K_4)$  compatible by modifying the form of $R_4$ and
$K_4$ as
\begin{eqnarray}
\hat{R}  = F(t,x,y,\dot{x},\dot{y})R,\qquad \hat{K}  =
G(t,x,y,\dot{x},\dot{y})K, \label{tc03}
\end{eqnarray}
where $F$ and $G$ are functions to be determined. However, after
making these modifications, $S_4$ and $U_4$ remain the same. A
motivation to do this type of modification came from our earlier
work on the applicability of the PS procedure to scalar second order
ODEs (Chandrasekar \textit{et al.} 2005$a$) . However, unlike the
scalar case, now we have to choose two functions, namely $F$ and
$G$, appropriately, such that the compatible forms of $\hat{R}$ and
$\hat{K}$ can be fixed. Since we have to choose two functions one
might ask the question whether these two  functions are necessarily
different or the same. In the following we answer this question.

Since $\hat{R}$ and $\hat{K}$ should
satisfy equations~(\ref{eq25}) and (\ref{eq26}), we have
\begin{align}
D[F]R+FD[R]=-(FR\phi_{1\dot{x}}+GK\phi_{2\dot{x}}+FRS),
\label {tc04}\\
D[G]K+GD[K]=-(FR\phi_{1\dot{y}}+GK\phi_{2\dot{y}}+GKU).
\label {tc05}
\end{align}
Substituting the expressions for $D[R]$ and $D[K]$ (vide (\ref{eq25}) and
(\ref{eq26})) in (\ref{tc04}) and (\ref{tc05}) and
simplifying the resultant equations we find that
\begin{align}
D[F]R=(F-G)K\phi_{2\dot{x}}, \quad D[G]K=(G-F)R\phi_{1\dot{y}}.
\label {tc06}
\end{align}
On the other hand substituting the modified forms (\ref{tc03}) in the
integrability conditions (\ref{eq33}), we get
\begin{align}
D[F]RS=(F-G)K\phi_{2x}, \quad D[G]KU=(G-F)R\phi_{1y}.
\label {tc07}
\end{align}
Combining equations (\ref{tc06}) and (\ref{tc07}) we obtain
\begin{align}
(F-G)K(S\phi_{2\dot{x}}-\phi_{2x})=0, \quad
(G-F)R(U\phi_{1\dot{y}}-\phi_{1y})=0.
\label {tc07a}
\end{align}
From equation (\ref{tc07a}) we can conclude that either
$F\neq G$ or $F=G$. In the first case, we have uncoupled equations, that is,
\begin{align}
F\neq G \Rightarrow
\phi_{2\dot{x}}=\phi_{1\dot{y}}=\phi_{2x}=\phi_{1y}=0\quad
(\hat{R}=FR,\;\hat{K}=GK)
\label {tc07aa}
\end{align}
 and in the second case we have coupled equations, that is,
\begin{align}
F= G \Rightarrow
\phi_{2\dot{x}},\phi_{1\dot{y}},\phi_{2x},
\phi_{1y}\neq0\quad
(\hat{R}=FR,\;\hat{K}=FK).
\label {tc07ab}
\end{align}
Note that in the case of $F\neq G$, the other possible solution
$S=\phi_{2x}/\phi_{2\dot{x}}$, $U=\phi_{1y}/\phi_{1\dot{y}}$, when
used in (\ref{eq23}) and (\ref{eq24}) lead to inconsistencies and so
this choice is not considered. Further, the case $R=0$ and $K=0$ leads to 
the trivial solution. The above analysis clearly shows that
for the uncoupled equations one should choose $F$ and $G$ as
different and for the coupled equations one should choose them as
the same.

Finally, we mention the important point that the functions $F$ and $G$
are nothing but functions of integrals of motion. To show this,
substituting back the relations (\ref{tc07aa}) and (\ref{tc07ab}) in
(\ref{tc06}) and (\ref{tc07}), respectively, we get
\begin{align}
D[F]R=0,\; D[G]K=0 \; \mbox{and}\; D[F]RS=0,\; D[G]KU=0,\;
\Rightarrow D[F]=0= D[G].\label {tc06b}
\end{align}
Thus, irrespective of whether the given equations are coupled ones
or uncoupled ones the functions $F$ and $G$ can always be taken as
integrals of motion or functions of them and the problem now is how
to determine them. In the following, we discuss this in detail.

\section{Case 1: $F\neq G$ (Uncoupled Equations) }
\subsection{Theory}
In this case, we have the following form of equation of motion,
namely
\begin{eqnarray}
\ddot{x}=\phi_1(t,x,\dot{x}), \quad
\ddot{y}=\phi_2(t,y,\dot{y}).\label {tc08}
\end{eqnarray}
As a consequence the evolution equations for the functions $S,U,R$ and $K$ (vide equations (\ref{eq23})-(\ref{eq26})) are simplified to the forms
\begin{eqnarray}
D[S] &=&-\phi_{1x}+S\phi_{1\dot{x}}+S^2,  \label {eq23a}\\
D[U] &=&-\phi_{2y}+U\phi_{2\dot{y}}+U^2,  \label {eq24a}\\
D[R]  &=&-R S-R\phi_{1\dot{x}},  \label {eq25a}\\
D[K]  &=&-K U-K\phi_{2\dot{y}}. \label {eq26a}
\end{eqnarray}
Solving equations (\ref{eq23a}) and (\ref{eq24a}) one can get explicit
forms of $S$ and $U$. Substituting the known forms of $S$ and $U$ into the
equations (\ref{eq25a}) and (\ref{eq26a}) and integrating the resultant
equations one can fix the forms of $R$ and $K$. From the known forms of
$(S,U,R,K)$ one can fix the integrals of motion from (\ref{cso09}). 
One can also consider equation (\ref{tc08}) as two independent 
second order ODEs and solve the equations independently 
by adopting the procedure given in our previous paper 
(Chandrasekar \textit{et al.} 2005$a$). To avoid repetition we do 
not discuss the uncoupled case further here.

\section{Case 2: $F=G$ (Coupled equations)}
\subsection{Theory}
Now we focus on the case $F=G$. To obtain the forms $(S,U,R,K)$ one
needs to solve the compatibility conditions (\ref{eq23}) -
(\ref{eq26}). For this purpose, we rewrite
equations~(\ref{eq23})-(\ref{eq26}) in terms of two variables alone, namely $R$
and $K$, by eliminating $S$ and $U$, and analyse the resulting
coupled second order partial differential equations (PDEs) and obtain expressions for $R$ and $K$.
From the latter we deduce the forms of $S$ and $U$ through the
relations (\ref{eq25}) and (\ref{eq26}).

To rewrite equations~(\ref{eq23})-(\ref{eq26}) interms of $R$ and $K$ let us
take a total derivative of equations (\ref{eq25}) and (\ref{eq26}). Doing so we get
\begin{align}
D^2[R]=-D[R\phi_{1\dot{x}}+K\phi_{2\dot{x}}+RS],\quad
D^2[K]=-D[K\phi_{2\dot{y}}+R\phi_{1\dot{y}}+KU].
\label {3.1}
\end{align}
Using the identities (\ref{eq33}), equation (\ref{3.1}) can be
rewritten in a coupled form for $R$ and $K$ as
\begin{align}
D^2[R]+D[R\phi_{1\dot{x}}+K\phi_{2\dot{x}}]=R\phi_{1x}+K\phi_{2x},
\label {tc01}\\
D^2[K]+D[K\phi_{2\dot{y}}+R\phi_{1\dot{y}}]=R\phi_{1y}+K\phi_{2y}.
\label {tc02}
\end{align}
One can note that the above determining equations (\ref{tc01}) and
(\ref{tc02}) form a system of linear PDEs. To solve 
equations~(\ref{tc01}) and (\ref{tc02}) one may assume a specific ansatz either polynomial or rational in $\dot{x}$ and $\dot{y}$ for $R$ and $K$ and substituting 
the known expressions of $\phi_1$ and $\phi_2$ and their derivatives 
into (\ref{tc01}) and
(\ref{tc02}) and solving them one can obtain expressions for the
integrating factors $R$ and $K$. Once $R$ and $K$ are known then the
functions $(S,U)$ can be fixed through the relation
(\ref{eq25})-(\ref{eq26}). Knowing $S,U,R$ and $K$, one has to make
sure that this set $(S,U,R,K)$ does also satisfy the remaining
compatibility conditions (\ref{eq31}) - (\ref{eq32}). The set
$(S,U,R,K)$ which satisfies all the equations (\ref{eq23}) -
(\ref{eq32}) is then the acceptable solution and one can then
determine the associated integral $I$ using the relation
(\ref{cso09}). For complete integrability we require four
independent compatible sets $(S_i,U_i,R_i,K_i),\;i=1,2,3,4$.

As discussed in \S3, suppose the sets
$(S_i,U_i,R_i,K_i),\;i=1,2,3,$ are found to satisfy the
equations~(\ref{eq23}) - (\ref{eq32}) and the fourth set $(S_4,U_4,$
$R_4,K_4)$ does not satisfy equations (\ref{eq31}) - (\ref{eq32}).
In this case to identify the correct form of $\hat{R_4}$ and
$\hat{K_4}$ one may assume that $\hat{R_4} = F(I_1,I_2,I_3)R_4$ and
$\hat{K_4} = F(I_1,I_2,I_3)K_4$, where $F(I_1,I_2,I_3)$ is a
function of the integrals $I_1,\;I_2$ and $I_3$. To determine the
explicit form of $F(I_1,I_2,I_3)$ we proceed as follows.
Substituting
\begin{eqnarray}
\hat{R_4}  = F(I_1,I_2,I_3)R_4 ,\quad \hat{K_4}  = F(I_1,I_2,I_3)K_4
\label{ext06}
\end{eqnarray}
into equations~(\ref{eq31})-(\ref{eq32}), we get the following 
set of six relations:
\begin{align}
\frac{(a_1F_1'+b_1F_2'+c_1F_3')}{d_1}=F,\quad
\frac{(a_2F_1'+b_2F_2'+c_2F_3')}{d_2}=F,\nonumber\\
\frac{(a_3F_1'+b_{3}F_2'+c_{3}F_3')}{d_{3}}=F,\quad
\frac{(a_4F_1'+b_{4}F_2'+c_{4}F_3')}{d_{4}}=F,\nonumber\\
\frac{(a_5F_1'+b_{5}F_2'+c_{5}F_3')}{d_{5}}=F,\quad
\frac{(a_6F_1'+b_{6}F_2'+c_{6}F_3')}{d_{6}}=F,
\label{meth02}
\end{align}
where
\begin{eqnarray}
&&a_1=R(I_{1x}-SI_{1\dot{x}}),\quad
b_1=R(I_{2x}-SI_{2\dot{x}}),\quad
c_1=R(I_{3x}-SI_{3\dot{x}}),
\nonumber\\
&&d_1=(SR_{\dot{x}}+RS_{\dot{x}}-R_{x}),\quad
a_2=K(I_{1y}-UI_{1\dot{y}}),\quad
b_2=K(I_{2y}-UI_{2\dot{y}}),
\nonumber\\
&&c_2=K(I_{3y}-UI_{3\dot{y}}),\quad
d_2=(UK_{\dot{y}}+KU_{\dot{y}}-K_{y}),\quad
a_3=(RI_{1y}-UKI_{1\dot{x}}),
\nonumber\\
&&b_{3}=(RI_{2y}-UKI_{2\dot{x}}),\quad
c_{3}=(RI_{3y}-UKI_{3\dot{x}}),\quad
d_{3}=(UK_{\dot{x}}+KU_{\dot{x}}-R_{y}),
\nonumber\\
&&a_{4}=(KI_{1x}-SRI_{1\dot{y}}),\quad
b_{4}=(KI_{2x}-SRI_{2\dot{y}}),\quad
c_{4}=(KI_{3x}-SRI_{3\dot{y}}),
\nonumber\\
&&d_{4}=(SR_{\dot{y}}+RS_{\dot{y}}-K_{x}),\quad
a_{5}=(RI_{1\dot{y}}-KI_{1\dot{x}}),\quad
b_{5}=(RI_{2\dot{y}}-KI_{2\dot{x}}),
\nonumber\\
&&c_{5}=(RI_{3\dot{y}}-KI_{3\dot{x}}),\quad
d_{5}=(K_{\dot{x}}-R_{\dot{y}}),\quad
a_{6}=(SRI_{1y}-UKI_{1x}),
\nonumber\\
&&b_{6}=(SRI_{2y}-UKI_{2x}),\quad
c_{6}=(SRI_{3y}-UKI_{3x}),
\nonumber\\
&&d_{6}=(UK_{x}+KU_{x}-SR_{y}-RS_{y})\nonumber
\label{meth03}
\end{eqnarray}
are all known functions of $t,x,y,\dot{x}$ and $\dot{y}$ and
$F_{i}'=\partial F/\partial I_i$.

Equation~(\ref{meth02}) represents an over determined  system of
equations for the unknown $F$. A simple way to solve this equation
is to uncouple it for $F_i'$, ($=\frac{\partial F}{\partial I_i}$)
$i=1,2,3,$ and solve the resultant equations. For example,
eliminating $F_2'$ and $F_3'$ from the first three relations in
equation~(\ref{meth02}) we obtain an equation for $F_1'$ in the form
\begin{align}
\frac{F_1'}{F}=\frac{(d_1c_2-c_1d_2)(b_1c_{3}-b_{3}c_1)
-(d_1c_3-c_1d_3)(b_1c_{2}-b_{2}c_1)}{(a_1c_2-c_1a_2)(b_1c_{3}-b_{3}c_1)
-(a_1c_3-c_1a_3)(b_1c_{2}-b_{2}c_1)}.
\label{meth04}
\end{align}
On the other hand eliminating $F_1'$ and $F_3'$ from equation~(\ref{meth02})
(again from the first three relations)
we arrive at the following equations for $F_2'$ in the form
\begin{align}
\frac{F_2'}{F}=\frac{(d_1c_2-c_1d_2)(a_1c_{3}-a_{3}c_1)
-(d_1c_3-c_1d_3)(a_1c_{2}-a_{2}c_1)}{(b_1c_2-c_1b_2)(a_1c_{3}-a_{3}c_1)
-(b_1c_3-c_1b_3)(a_1c_{2}-a_{2}c_1)}.
\label{meth05}
\end{align}
In the similar way one obtains the following equation for $F_3'$
\begin{align}
\frac{F_3'}{F}=\frac{(d_1b_2-b_1d_2)(a_1b_{3}-a_{3}b_1)
-(d_1b_3-b_1d_3)(a_1b_{2}-a_{2}b_1)}{(c_1b_2-b_1c_2)(a_1b_{3}-a_{3}b_1)
-(c_1b_3-b_1c_3)(a_1b_{2}-a_{2}b_1)}.
\label{meth06}
\end{align}

One can easily check that the combination of other relations in equation
(\ref{meth02}) along with the forms (\ref{meth04})-(\ref{meth06}) gives rise to
relations which are effectively nothing but the constraint
equations~(\ref{eq31})-(\ref{eq32}) and so no new constraint actually arises.
Consequently, equations (\ref{meth04})-(\ref{meth06}) can be written as
\begin{align}
\frac{\partial F}{\partial I_1}=g_1(I_1,I_2,I_3)F,\quad
\frac{\partial F}{\partial I_2}=g_2(I_1,I_2,I_3)F \quad  \mbox{and}\quad
\frac{\partial F}{\partial I_3}=g_3(I_1,I_2,I_3)F,
\label{meth07}
\end{align}
where $g_i$'s, $i=1,2,3,$ are functions of $I_1$, $I_2$ and $I_3$.
Now solving equations (\ref{meth07}) one can obtain the
explicit form of $F(I_1,I_2,I_3)$.  Once $F$ is known we can obtain the
complete solution $\hat{R_4}$ and $\hat{K_4}$ from which, along with $S_4$ and
$U_4$, the fourth integral $I_4$ can be constructed using the expression
(\ref{cso09}).

Finally, we note that in some cases one can meet the situation that
the sets $(S_i,U_i,R_i,K_i),\;i=1,2,$ alone are found to satisfy the
equations~(\ref{eq23}) - (\ref{eq32}) and the third set
$(S_3,U_3,R_3,K_3)$ (as well as the fourth set) does not satisfy
equations (\ref{eq31}) - (\ref{eq32}). In this case $F$ may be a
function of the integrals $I_1$ and $I_2$ which can be derived from
the sets $(S_i,U_i,R_i,K_i),\;i=1,2$. We need to find the explicit
form of $F(I_1,I_2)$ in order to obtain the compatible solution
$(S_3,U_3,\hat{R_3},\hat{K_3})$. To recover the complete form of
$\hat{R_3}$ and $\hat{K_3}$ one may assume that $\hat{R_3} =
F(I_1,I_2)R_3$ and $\hat{K_3} = F(I_1,I_2)K_3$ and proceed as before 
and obtain the determining equations for $F$. Since $F$ is a function 
of $I_1$ and $I_2$ alone one essentially gets the same form (\ref{meth02}) 
but without the factor $F_3'$ ($=\frac{\partial F}{\partial I_3}$). 
Since $c_i$'s $i=1,...6$, are also exclusive functions of $I_3$ and 
their derivatives they do not appear in the determining equations. 
Solving the 
resultant determining equations one can fix the form of $F$ which in turn 
provides us $\hat{R}_3$ and $\hat{K}_3$ from which one can 
constract the third integral $I_3$ for the given problem. 
Now from the known
forms $I_1,I_2$ and $I_3$, one can proceed as before to obtain the
fourth compatible set $(S_4,U_4,\hat{R}_4,\hat{K}_4)$ and thereby
obtain the fourth integral $I_4$ also.

Similarly, if the set $(S_1,U_1,R_1,K_1)$ alone is found to satisfy
the equations~(\ref{eq23}) - (\ref{eq32}) and the second set
$(S_2,U_2,R_2,K_2)$ does not satisfy equations (\ref{eq31}) -
(\ref{eq32}) then, in this case the determining equations for $F$
takes the form (\ref{meth02}) with $F_2=F_3=b_i=c_i=0,\;i=1,...6$.
Following the procedure mentioned above one can first derive 
the second integral. From a knowledge of $(I_1,I_2)$, one can
again follow the above procedures and construct $I_3$ and $I_4$.
In the following we illustrate the theory with specific examples.

\subsection{Example 2: Coupled modified Emden equation}
The modified Emden equation (MEE) and its variants arise in different branches of
physics (Erwin \textit{et al.} 1984; Dixon \& Tuszynski 1990) and
considerable attention has  been paid recently to explore the mathematical
and geometrical properties of MEE and its variant equations (see for example,
Mahomed \& Leach 1985; Steeb 1993; Chandrasekar
\textit{et al.} 2005$b$; Euler \textit{et al.} 2007). The equation reads
\begin{eqnarray}
\ddot{x}+3kx\dot{x}+kx^3+\lambda x=0,
\label {MEE01}
\end{eqnarray}
where $k$ and $\lambda$ are arbitrary parameters,
which can be completely integrated in terms of elementary functions
(Chandrasekar \textit{et al.} 2005$b$). 
It exhibits some unusual features such as amplitude independent frequency, 
nonstandard Lagrangian and time independent Hamiltonian forms and 
transition from periodic to front-like solutions as the sign of the 
parameter changes (Chandrasekar \textit{et al.} 2005$b$).

We now consider a coupled two dimensional version of (\ref{MEE01})
and demonstrate how it can be integrated out by following the theory
given in the previous section. The coupled system of equations reads
\begin{eqnarray}
&&\ddot{x}+2(k_1x+k_2y)\dot{x}+(k_1\dot{x}+k_2\dot{y})x+(k_1x+k_2y)^2x
+\lambda_1 x=0,\nonumber\\
&&\ddot{y}+2(k_1x+k_2y)\dot{y}+(k_1\dot{x}+k_2\dot{y})y+(k_1x+k_2y)^2y
+\lambda_2 y=0,
\label {case32}
\end{eqnarray}
where $k_i$ and $\lambda_i$, $i=1,2$, are arbitrary parameters.

To determine the integrating factors we seek a rational form of ansatz for
$R$ and $K$ in the form (suggested by the equation (\ref{MEE01}))
\begin{align}
R=\frac{a_1+a_2\dot{x}+a_3\dot{y}}
{(a_4+a_5\dot{x}+a_6\dot{y})^q},\qquad K=\frac{b_1+b_2\dot{x}+b_3\dot{y}}
{(a_4+a_5\dot{x}+a_6\dot{y})^r}
 \label{case32a},
\end{align}
where $q$ and $r$ are arbitrary numbers (to be fixed) and $a_i$'s, $i=1,2...6$
and $b_i$'s, $i=1,2,3,$ are arbitrary functions of $t,\;x$ and $y$. Substituting
(\ref{case32a}) into (\ref{tc01}) and (\ref{tc02}) and equating the
coefficients of different powers of $\dot{x}$ and $\dot{y}$ to zero we get a
set of linear PDEs for the variables $a_i$'s,
$i=1,2...6,$ and $b_j$'s, $j=1,2,3$. Solving the resultant equations we find the
following sets of particular solutions for $R$ and $K$:
\begin{eqnarray}
&&R_1=-\frac{e^{-\sqrt{-\lambda_1}t}(\lambda_1(k_2\dot{y}+k_2^2y^2+\lambda_2)
+k_1x(k_2\lambda_1y-\lambda_2\sqrt{-\lambda_1}))}
{g(x,y,\dot{x},\dot{y})^2}, \nonumber\\
&&K_1=\frac{e^{-\sqrt{-\lambda_1}t}\lambda_1k_2(\dot{x}+k_1x^2+x(\sqrt{-\lambda_1}+k_2y))}
{g(x,y,\dot{x},\dot{y})^2},\label{case33a}\\
&&R_2=\frac{e^{-\sqrt{-\lambda_2}t}\lambda_2k_1(\dot{y}+k_2y^2+y(\sqrt{-\lambda_2}+k_1x))}
{g(x,y,\dot{x},\dot{y})^2},\nonumber\\
&&K_2=-\frac{e^{-\sqrt{-\lambda_2}t}(\lambda_2(k_1\dot{x}+k_1^2x^2+\lambda_1)
+k_2y(\lambda_2k_1x-\lambda_1\sqrt{-\lambda_2}))}
{g(x,y,\dot{x},\dot{y})^2},\label{case33b}\\
&&R_3=\frac{x}
{g(x,y,\dot{x},\dot{y})^2}, \quad K_3=0,\label{case33c}\\
&&R_4=0, \quad
K_4=\frac{y}
{g(x,y,\dot{x},\dot{y})^2},
\label{case33d}
\end{eqnarray}
where $g(x,y,\dot{x},\dot{y})=(k_1\lambda_2\dot{x}+k_2\lambda_1\dot{y}+k_1^2\lambda_2x^2
+k_2^2\lambda_1y^2+k_1k_2(\lambda_1+\lambda_2)xy+\lambda_1\lambda_2)$. Now substituting the above forms of $R_i$'s
and $K_i$'s, $i=1,2,3,4,$  into (\ref{eq25}) and (\ref{eq26}), we
can obtain the corresponding $S_i$'s and $U_i$'s, $i=1,2,3,4$. As a
consequence now we have four sets of independent solutions
$(S_i,U_i,R_i, K_i)$, $i=1,2,3,4,$ for the equations
(\ref{eq23})-(\ref{eq26}). Now we check the compatibility of these
solutions with the remaining equations (\ref{eq31})-(\ref{eq32}). We
find that only the first two sets $(S_i,U_i,R_i, K_i)$, $i=1,2,$
satisfy the extra constraints (\ref{eq31})-(\ref{eq32}), and become
compatible solutions. Substituting their forms separately into
equation (\ref{cso09}) and evaluating the integrals we get
\begin{eqnarray}
I_1=\frac{e^{-\sqrt{-\lambda_1}t}(\dot{x}+(k_1x+k_2y)x+\sqrt{-\lambda_1}x)}
{g(x,y,\dot{x},\dot{y})},\nonumber\\
I_2=\frac{e^{-\sqrt{-\lambda_2}t}(\dot{y}+(k_1x+k_2y)y+\sqrt{-\lambda_2}y)}
{g(x,y,\dot{x},\dot{y})}.
\label{case34}
\end{eqnarray}
However, the sets $(S_i,U_i,R_i, K_i)$'s, $i=3,4,$ do not satisfy
the extra constraints (\ref{eq31})-(\ref{eq32}) which means that the
form of $R_3$ in the third set and $K_4$ in the fourth set may not
be in  the 'complete form' (since $K_3=R_4=0$)
but might only be a factor of the
complete form.

To deduce the compatible set $(S_3,\;U_3,\;\hat{R}_3,\;K_3)$, let us substitute
$\hat{R_3}=F(I_1,I_2)$ $R_3$ into equations~(\ref{meth04}) - (\ref{meth06}).
As a result we get
\begin{eqnarray}
\frac{1}{2}I_1F_1'+F = 0,\quad F_2'=0,\quad F_3'=0, \qquad
(F_i=\frac{\partial F}{\partial I_i},\; i=1,2,3).
\label{case34b}
\end{eqnarray}
Upon integrating (\ref{case34b}), we get $F = 1/I_1^2$,
(the integration constants are
set to zero for simplicity) which fixes the form of $\hat{R_3}$ as
\begin{eqnarray}
\hat{R}_3=\frac{e^{2\sqrt{-\lambda_1}t}x}
{(\dot{x}+(k_1x+k_2y)x+\sqrt{-\lambda_1}x)^2}.
\label{case35}
\end{eqnarray}
Now one can easily check that the set $(S_3,U_3,\hat{R}_3,K_3)$ is a
compatible solution for the full set of determining
equations~(\ref{eq23})-(\ref{eq32}) which in turn provides $I_3$
through the relation (\ref{cso09}) as
\begin{eqnarray}
I_3=\frac{e^{2\sqrt{-\lambda_1}t}(\dot{x}+(k_1x+k_2y)x-\sqrt{-\lambda_1}x)}
{(\dot{x}+(k_1x+k_2y)x+\sqrt{-\lambda_1}x)}.
\label{case36}
\end{eqnarray}

Finally, in the fourth set we take $\hat{K}_4=F(I_1,I_2,I_3)K_4$ and
substitute it into equations~(\ref{meth04})-(\ref{meth06}), to obtain
\begin{eqnarray}
F_1'= 0,\quad \frac{1}{2}I_2F_2'+F=0,\quad F_3'=0.
\label{case34c}
\end{eqnarray}
Upon integrating (\ref{case34c}), we get $F = 1/I_2^2$
which fixes the form of $\hat{K_4}$ as
\begin{eqnarray}
\hat{K}_4=\frac{e^{2\sqrt{-\lambda_2}t}y}
{(\dot{y}+(k_1x+k_2y)y+\sqrt{-\lambda_2}y)^2}.
\label{case35a}
\end{eqnarray}
Now the set $(S_4,U_4,R_4,\hat{K}_4)$ is a compatible solution for the
equations~(\ref{eq23})-(\ref{eq32}) which in
turn provides $I_4$ through the relation (\ref{cso09}) as
\begin{eqnarray}
I_4=\frac{e^{2\sqrt{-\lambda_2}t}(\dot{y}+(k_1x+k_2y)y-\sqrt{-\lambda_2}y)}
{(\dot{y}+(k_1x+k_2y)y+\sqrt{-\lambda_2}y)}.
\label{case36a}
\end{eqnarray}
Using the explicit forms of the integrals, $I_1,\; I_2,\;I_3$ and $I_4$, the
general solution to equation~(\ref{case32}) can be deduced directly as
\begin{eqnarray}
x(t)=\frac{\sqrt{-\lambda_1}\lambda_2 I_1(e^{2\sqrt{-\lambda_1}t}-I_3)}
{(\hat{I}_1+I_2k_2\lambda_1e^{\hat{\lambda}_1t}
+I_2I_4k_2\lambda_1e^{\hat{\lambda}_2t}
+I_1k_1\lambda_2e^{2\sqrt{-\lambda_1}t}
-2e^{\sqrt{-\lambda_1}t})},\;\;
\nonumber\\
y(t)=\frac{\sqrt{-\lambda_2}\lambda_1 I_2(e^{2\sqrt{-\lambda_2}t}-I_4)}
{(\hat{I}_2+I_1k_1\lambda_2e^{\hat{\lambda}_1t}
+I_1I_3k_1\lambda_2e^{-\hat{\lambda}_2t}
+I_2k_2\lambda_1e^{2\sqrt{-\lambda_2}t}
-2e^{\sqrt{-\lambda_2}t})},\label {case37}
\end{eqnarray}
where $\hat{\lambda}_1=\sqrt{-\lambda_1}+\sqrt{-\lambda_2},
\;\hat{\lambda}_2=\sqrt{-\lambda_1}-\sqrt{-\lambda_2},\;\hat{I}_1=I_1I_3k_1\lambda_2$ and $\hat{I}_2=I_2I_4k_2\lambda_1$.

Finally we note that for the case $\lambda_i>0$, $i=1,2,$ the above general
solution becomes a periodic one, that is
\begin{eqnarray}
&&x(t)=\frac{A\sin(\omega_1 t+\delta_1)}
{1-\frac{Ak_1}{\omega_1}\cos(\omega_1 t+\delta_1)
-\frac{Bk_2}{\omega_2}\cos(\omega_2 t+\delta_2)},\label {cmee04a}\\
&&y(t)=\frac{B\sin(\omega_2 t+\delta_2)}
{1-\frac{Ak_1}{\omega_1}\cos(\omega_1 t+\delta_1)
-\frac{Bk_2}{\omega_2}\cos(\omega_2 t+\delta_2)},\label {cmee04b}
\end{eqnarray}
where $\omega_j=\sqrt{\lambda_j},\;j=1,2,\; A=\omega_1\omega_2^2 I_1e^{-i\delta_1},
\;B=\omega_2\omega_1^2I_2e^{-i\delta_2},\;I_3=e^{-2i\delta_1}$ and $I_4=e^{-2i\delta_2}$
are arbitrary constants and $(Ak_1/\omega_1+Bk_2/\omega_2)<1$. More details on
the classical dynamics and characteristic features of this system will be
published elsewhere.

\subsection{Example 3: Coupled Mathews - Lakshmanan oscillator}
Let us consider  the coupled Mathews - Lakshmanan oscillator discussed
by Cari$\tilde{n}$ena \textit{et al.} (2004$a$;2004$b$;2007),  namely
\begin{eqnarray}
&&\ddot{x}=\frac{\lambda(\dot{x}^2+\dot{y}^2
+\lambda(x\dot{y}-y\dot{x})^2)x-\alpha^2x}{(1+\lambda r^2)}=\phi_1,\nonumber\\
&&\ddot{y}=\frac{\lambda(\dot{x}^2+\dot{y}^2
+\lambda(x\dot{y}-y\dot{x})^2)y-\alpha^2y}{(1+\lambda r^2)}=\phi_2,
\label{ext01}
\end{eqnarray}
where $r=\sqrt{x^2+y^2}$ and $\lambda$ and $\alpha$ are arbitrary parameters,
Equation (\ref{ext01}) is a two dimensional generalization of the one
dimensional non-polynomial oscillator introduced by Mathews \& Lakshmanan (1974),
\begin{eqnarray}
(1+\lambda x^2)\ddot{x}-(\lambda\dot{x}^2-\alpha^2)x=0.
\label{sin01}
\end{eqnarray}

Now we use the following ansatz for $R$ and $K$ to explore the
integrating factors for the equation (\ref{ext01}) in the form
(suggested by the one dimensional oscillator (\ref{sin01}))
\begin{eqnarray}
R&=&a_1+a_2\dot{x}+a_3\dot{y}+a_4\dot{x}^2
+a_5\dot{x}\dot{y}+a_6\dot{y}^2+a_7\dot{x}^3+a_8\dot{x}^2\dot{y}
+a_9\dot{y}^2\dot{x}+a_{10}\dot{y}^3,\nonumber\\
K&=&b_1+b_2\dot{x}+b_3\dot{y}+b_4\dot{x}^2
+b_5\dot{x}\dot{y}+b_6\dot{y}^2+b_7\dot{x}^3+b_8\dot{x}^2\dot{y}
+b_9\dot{y}^2\dot{x}+b_{10}\dot{y}^3,\nonumber\\
\label {anz01}
\end{eqnarray}
where $a_i$'s and $b_i$'s $i=1,2...10,$ are arbitrary functions of $t,\;x$ and $y$. Substituting (\ref{anz01}) into (\ref{tc01}) and (\ref{tc02})
and equating the coefficients
of different powers of $\dot{x}$ and $\dot{y}$ to zero and solving 
the resultant PDEs as before
one obtains the following nontrivial solutions for $R$ and $K$:
\begin{eqnarray}
&&R_1=y, \quad K_1=-x,\\
&&R_2=\frac{2\lambda((1+\lambda y^2)\dot{x}-\lambda xy\dot{y})}
{1+\lambda r^2}, \quad
K_2=\frac{2\lambda((1+\lambda x^2)\dot{y}-\lambda xy\dot{x})}
{1+\lambda r^2},\\
&&R_3=\dot{x}+\dot{y}
+\frac{\lambda (x+y)^2(\lambda xy\dot{y}-(1+\lambda y^2)\dot{x})}
{1+\lambda r^2}, \nonumber\\
&&K_3=\dot{x}+\dot{y}
-\frac{\lambda (x+y)^2((1+\lambda x^2)\dot{y}-\lambda xy\dot{x})}
{1+\lambda r^2},\\
&&R_4=\frac{-1}{1+\lambda r^2}\bigg[t\lambda \bigg((\dot{x}+\dot{y})^2
-\frac{(x+y)^2\phi_1}{x}\bigg)\bigg((1+\lambda y^2)\dot{x}-\lambda xy\dot{y}\bigg)
\nonumber\\&&\qquad
-(x+y)(\alpha^2+\lambda(1+\lambda y(x+y))\dot{x}\dot{y}
-\lambda(1+\lambda x(x+y))\dot{y}^2)\bigg],\\
&&K_4=\frac{-1}{1+\lambda r^2}\bigg[t\lambda \bigg((\dot{x}+\dot{y})^2
-\frac{(x+y)^2\phi_2}{y}\bigg)\bigg((1+\lambda x^2)\dot{y}-\lambda xy\dot{x}\bigg)
\nonumber\\&&\qquad
-(x+y)(\alpha^2-\lambda y(1+\lambda(x+y))\dot{x}^2
+\lambda(1+\lambda x(x+y))\dot{x}\dot{y})\bigg].
\label{ext02}
\end{eqnarray}

Now substituting the above forms of $R_i$'s and $K_i$'s, $i=1,2,3,4,$
into the equation (\ref{eq25}) and (\ref{eq26}), we get the corresponding forms of
$S_i$'s and
$U_i$'s, $i=1,2,3,4$. As a consequence now we have four sets of independent
solutions for the equation
(\ref{eq23})-(\ref{eq26}). Out of these we find that only the first three sets
($i=1,2,3$) satisfy the extra constraints (\ref{eq31})-(\ref{eq32}) and become
compatible solutions. Substituting these forms separately into equation
(\ref{cso09}) and evaluating the integrals we arrive at
\begin{align}
I_1& =(y\dot{x}-x\dot{y}), \qquad
I_2=\frac{(\alpha^2-\lambda(\dot{x}^2+\dot{y}^2+\lambda(y\dot{x}-x\dot{y})^2))}
{1+\lambda r^2},\label{ext04}\\
I_3&=(\dot{x}+\dot{y})^2+
\frac{(x+y)^2(\alpha^2-\lambda(\dot{x}^2+\dot{y}^2+\lambda(y\dot{x}-x\dot{y})^2))}
{1+\lambda r^2}.\label{ext05}
\end{align}

However, the set $(S_4,U_4,R_4, K_4)$ does not satisfy the extra
constraints (\ref{eq31})-(\ref{eq32}) which means that the forms of
$R_4$ and $K_4$ are incomplete. So we assume
that $\hat{R_4}$ and $\hat{K_4}$ are in the forms (\ref{ext06}). As
a consequence equations (\ref{meth04}) - (\ref{meth06}) lead us to
the following determining equations for the unknown $F$, namely
\begin{equation}
F_1'=0,\quad 2I_2F_2'+F = 0,\quad I_3F_3'+F=0, \quad
(F_i=\frac{\partial F}{\partial I_i},\; i=1,2,3).
\label{ext07}
\end{equation}
Upon integrating (\ref{ext07}) we get,
$F = \frac {1}{\sqrt{I_2}I_3}$, (the integration constants are set to zero
for simplicity) which fixes the form of $\hat{R_4}$ and $\hat{K_4}$ as
\begin{eqnarray}
\hat{R}_4=\frac{R_4}{\sqrt{I_2}I_3} \qquad
\hat{K}_4=\frac{K_4}{\sqrt{I_2}I_3},
\label{ext08}
\end{eqnarray}
where $I_2$ and $I_3$ are given in equations (\ref{ext04}) and (\ref{ext05}).
Now one can easily check that the set $(S_4,U_4,\hat{R}_4, \hat{K}_4)$
is a compatible solution for the equations~(\ref{eq23})-(\ref{eq32}) which in
turn provides $I_4$ through the relation (\ref{cso09}) in the form
\begin{eqnarray}
I_4&=&-\sqrt{\frac{t^2(\alpha^2-\lambda(\dot{x}^2+\dot{y}^2+\lambda(y\dot{x}
-x\dot{y})^2))}
{1+\lambda r^2}}\nonumber\\&&\qquad
+tan^{-1}\bigg[\sqrt{\frac{(x+y)^2(\alpha^2
-\lambda(\dot{x}^2+\dot{y}^2+\lambda(y\dot{x}-x\dot{y})^2))}
{(\dot{x}+\dot{y})^2(1+\lambda r^2)}}\bigg].
\label{ext09}
\end{eqnarray}
Note that the above integral is given for the first time in the literature.

Using the explicit forms of the integrals $I_1,\; I_2,\;I_3$ and $I_4$, the
solution to equation~(\ref{ext01}) can be deduced directly as
\begin{eqnarray}
x(t)=\sqrt{\frac{I_3}{4I_2}}\sin(\sqrt{I_2} t+I_4)
+\sqrt{\frac{\alpha^2-I_2-2\lambda I_3-\lambda^2 I_1^2}
{2\lambda I_2}}\sin(\sqrt{I_2} t+\delta),\nonumber\\
y(t)=\sqrt{\frac{I_3}{4I_2}}\sin(\sqrt{I_2} t+I_4)
-\sqrt{\frac{\alpha^2-I_2-2\lambda I_3-\lambda^2 I_1^2}
{2\lambda I_2}}\sin(\sqrt{I_2} t+\delta),
\label{ext10}
\end{eqnarray}
where $ \delta=I_4-\frac{1}{2}cos^{-1} [\frac{\lambda
I_1^2(4I_2+\lambda I_3) +I_3(7I_2-\alpha^2+2\lambda I_3)}
{I_3(I_2-\alpha^2+2\lambda I_3+\lambda^2 I_1^2)}]$.
\subsection{Example: 4 Known Two-dimensional Integrable Systems}
We have also studied the integrability properties of certain well
known two-dimensional nonlinear Hamiltonian systems, namely
Henon-Heiles system (Ramani \textit{et al.} 1989; Lakshmanan \&
Sahadevan 1993; Lakshmanan \& Rajasekar 2003) and generalized van
der Waals potential equation (Ganesan \& Lakshmanan 1990; 
Lakshmanan \& Rajasekar 2003) through the modified PS method. 
Our analysis shows that these systems
do admit only the known integrable cases. In the following we
present the salient features of our analysis.
\subsubsection{Henon-Heiles system}
Let us consider the generalized Henon-Heiles system (H\'enon \& Heiles 1964)
\begin{eqnarray}
\ddot{x}=-(Ax+2\alpha xy)\quad
\ddot{y}=-(By+\alpha x^2-\beta y^2)\label {h-heq01}
\end{eqnarray}
where $A,B,\alpha$ and $\beta$ are arbitrary parameters. We obtain
the integrating factors $R_1=\dot{x}$ and $K_1=\dot{y}$ for
arbitrary values of the parameters using the ansatz (\ref{anz01}).
Substituting the forms $R_1$ and $K_1$ into the equation
(\ref{eq25}) and (\ref{eq26}), we get the corresponding forms of
$S_1$ and $U_1$. Further, we find that this set also satisfies the extra
constraints (\ref{eq31})-(\ref{eq32}) and become a compatible
solution. Substituting these forms into the equation (\ref{cso09})
and evaluating the integrals we get
\begin{eqnarray}
I_1=\frac{1}{2}(\dot{x}^2+\dot{y}^2+Ax^2+By^2)+\alpha x^2y-\frac{\beta}{3} y^3,
\label {h-heq02}
\end{eqnarray}
which is nothing but the total energy. Further, we find that only
for the following specific parametric choices (i) $A=B,\;\alpha =-\beta$, (ii) $A,B\;
\mbox{arbitrary},\; \beta =-6\alpha$ and (iii) $B=16A,\; \beta
=-16\alpha$, one can get additional set of integrating
factors, that is,
\begin{eqnarray}
 (i)\quad &&R_2=-\dot{y},\qquad \qquad\qquad \qquad\qquad
\qquad\quad\; K_2=-\dot{x},\nonumber\\
 (ii) \quad && R_2=2\dot{x}y-x\dot{y}-\bigg(\frac{4A-B}{2\alpha}\bigg)\dot{x},
 \qquad \qquad K_2=-x\dot{x},\nonumber\\
 (iii)\quad && R_2=\alpha
 (x\dot{y}-6y\dot{x})x^2-3(\dot{x}^2+Ax^2)\dot{x},\quad K_2=\alpha\dot{x}x^3, \nonumber
\end{eqnarray}
respectively.

Finding the corresponding forms of $S$ and $U$ from (\ref{eq25}) and (\ref{eq26}) 
and substituting the above forms of $R$ and $K$ along with the corresponding 
$S$ and $U$ into (\ref{cso09}) and
evaluating the integrals we obtain the fol1owing second integrals 
for the above three parametric choices 
\begin{eqnarray}
(i)&&I_2=\dot{x}\dot{y}+(Ay+\alpha
(y^2+\frac{1}{3}x^2))x,\nonumber\\
(ii)&&I_2=4(x\dot{y}-\dot{x}y)\dot{x}+(4Ay+\alpha x^2+4\alpha
y^2)x^2
     +\frac{(4A-B)}{\alpha}(\dot{x}^2+Ax^2),\nonumber \label {h-heq03}\\
(iii)&&I_2=9(\dot{x}^2+Ax^2)^2+12\alpha\dot{x}x^2(3y\dot{x}-x\dot{y})
-2\alpha^2x^4(6y^2+x^2)-12\alpha Ax^4y,\nonumber   
\end{eqnarray}
respectively.

\subsubsection{Generalized van der Waals potential}
The generalized van der Waals potential equation in two dimensions
is given by
\begin{eqnarray}
\ddot{x}=-(2\gamma x+\frac{x}{r^{3}}),\quad
\ddot{y}=-(2\gamma \beta ^2y+\frac{y}{r^{3}}),\;\;r={(x^2+y^2)^{\frac {1}{2}}}
\label {vaneq01}
\end{eqnarray}
where $\gamma$ and $\beta$ are arbitrary parameters which can be
derived from the Hamiltonian
\begin{eqnarray}
H&=&\frac{1}{2}(\dot{x}^2+\dot{y}^2)+\gamma (x^2+\beta^2 y^2)
-\frac{1}{r}. \label {vaneq02}
\end{eqnarray}
Repeating the above procedure
we find that the system (\ref{vaneq01}) admits integrating factors
\begin{eqnarray}
 (i)\quad && R_2=-y,\qquad \quad\;\;\; K_2=x,\;\;\qquad\qquad \quad \beta^2=1\nonumber\\
 (ii) \quad && R_2=x\dot{y}-2y\dot{x},
 \quad K_2=x\dot{x},\qquad\quad\qquad \;
 \beta^2 =4,\nonumber\\
 (iii)\quad && R_2=y\dot{y},\qquad \qquad K_2=y\dot{x}-2x\dot{y},\qquad \; \beta^2 =\frac{1}{4}, 
\end{eqnarray}
and the corresponding second integrals of motion are (with $\gamma$ arbitrary)
\begin{eqnarray}
(i)&& I_2=(y\dot{x}-x\dot{y}),\qquad \qquad \qquad\;\;\;\;\beta^2=1,  \nonumber\\
(ii)&&I_2=(\dot{x}y-\dot{y}x)\dot{x}-\frac{y}{r}-2\gamma x^2y,\;\;
 \beta^2 =4,\nonumber\\
(iii)&&I_2=(\dot{y}x-\dot{x}y)\dot{y}-\frac{x}{r}- \frac{\gamma
y^2x}{2},\;\;\; \beta^2 =\frac{1}{4}. \label {vaneq03}
\end{eqnarray}

The integrals $I_1$ and $I_2$ ensure the integrability 
of the above systems for the respective parametric choices through 
Liouville theorem.

\section{Method of deriving general solution}
In the previous sections we derived the integrals of motion by deducing the
functions $S,\;U,\;R$ and $K$. However, in some situations it is very difficult
to solve the determining equations and obtain explicit forms for these
functions. Now we assume a situation in which we are able to derive only two
sets of solutions $(S_i,U_i,R_i,K_i),\;i=1,2$, and their associated integrals of motion and the remaining
two sets of functions are difficult to derive from the determining equations.
Under this situation, one may be able to deduce, in many cases, the remaining two integrals
from the known integrals themselves without
worrying further about the functions $(S_i, U_i, R_i,K_i)$, $i=3,4$. The  underlying idea is the following.

From the known integrals we deduce a set of co-ordinate
transformations and using these transformations we rewrite the
original ODEs as a set of two first order ODEs. By integrating the
latter we obtain the remaining two integration constants. The
solution to the original problem can then be obtained just by
inverting the variables. In the following, we first describe the
theory and then illustrate the ideas with an example.

We start the procedure with two known integrals  of motion, namely
\begin{eqnarray}
I_1=\mathcal{F}(t,x,y,\dot{x},\dot{y}),\quad \mbox{and}\quad
I_2=\mathcal{G}(t,x,y,\dot{x},\dot{y}).
\label{met13aa}
\end{eqnarray}

Now we split the functional form of the integrals $I_1$ and $I_2$
into two terms such that one involves all the variables
$(t,x,y,\dot{x},\dot{y})$ while
the other excludes $\dot{x}$ and $\dot{y}$, that is,
\begin{eqnarray}
I_1=F_1(t,x,y,\dot{x},\dot{y})+F_2(t,x,y),\quad
I_2=F_3(t,x,y,\dot{x},\dot{y})+F_4(t,x,y).
\label{met13a}
\end{eqnarray}
Let us split the functions $F_1$ and $F_3$ further in terms of two
functions, that is,
\begin{eqnarray}
&&I_1=F_1\left(\frac{1}{G_2(t,x,y,\dot{x},\dot{y})}\frac{d}{dt}G_1(t,x,y)\right)
+F_2\left(G_1(t,x,y)\right),\nonumber\\
&&I_2=F_3\left(\frac{1}{G_4(t,x,y,\dot{x},\dot{y})}\frac{d}{dt}G_3(t,x,y)\right)
+F_4\left(G_3(t,x,y)\right).
\label{met13b}
\end{eqnarray}

Now identifying a set of new variables in terms of functions $G_1$, $G_2$, $G_3$ and $G_4$ as
\begin{align}
w_1 = G_1(t,x,y),\quad z_1 = \int_o^t G_2(t',x,y,\dot{x},\dot{y}) dt',
\nonumber\\
w_2 = G_3(t,x,y),\quad z_2 = \int_o^t G_4(t',x,y,\dot{x},\dot{y}) dt',
\label{met13c}
\end{align}
equation (\ref{met13b}) can be rewritten in the form
\begin{eqnarray}
I_1=F_1\left(\frac {dw_1}{dz_1}\right)+F_2(w_1),\quad
I_2=F_3\left(\frac {dw_2}{dz_2}\right)+F_4(w_2).
\label{met13d}
\end{eqnarray}
In other words,
\begin{eqnarray}
F_1\left(\frac {dw_1}{dz_1}\right)=I_1-F_2(w_1),\quad
F_1\left(\frac {dw_2}{dz_2}\right)=I_2-F_2(w_2).
\label{met13e}
\end{eqnarray}
Rewriting equation (\ref{met13e}) one arrives at the decoupled equations
\begin{eqnarray}
\frac {dw_1}{dz_1}=f_1(w_1),\quad \frac {dw_2}{dz_2}=f_2(w_2),
\label{met13f}
\end{eqnarray}
which can in principle be integrated and two additional integration
constants identified. Now rewriting the solution in terms of the
original variables one obtains the general solution for the given
coupled second order ODE. Thus the new variables $w_1,\;w_2,\;z_1$ and $z_2$ correspond to transformations which effectively decouple the original coupled nonlinear second order ODEs (\ref{cso01}), provided such transformation variables can be identified. In the following we illustrate the
procedure with an example.

\subsection{Example:5 Force-free coupled Duffing-van der Pol oscillators}
Let us consider the force-free coupled Duffing-van der Pol
oscillators of the form
\begin{eqnarray}
\ddot{x}+(\alpha+\beta (x+y)^2)\dot{x}-\omega_1 x +\delta_1 x^3+\gamma_1
xy^2+\lambda_1 x^2y=0,\nonumber\\
\ddot{y}+(\alpha+\beta (x+y)^2)\dot{y}-\omega_2 y +\delta_2 y^3+\gamma_2
x^2y+\lambda_2 xy^2=0,
\label{cdvp01}
\end{eqnarray}
where  $\alpha$, $\beta,\;\omega_i,\;\delta_i,\;\gamma_i$ and $\lambda_i$,
$i=1,2,$ are arbitrary parameters.  Though the general equation (\ref{cdvp01})
has not been discussed in detail in the literature, special cases of the above 
coupled Duffing-van der
Pol equation have been used to represent physical and biological systems (Linkens 1974;
Datardina \& Linkens 1978; Kawahara 1980; Rajasekar \& Murali 2004).

We seek a rational form of ansatz for $R$ and $K$ in the form given
by (\ref{case32a}). Upon solving the
equations~(\ref{tc01})-(\ref{tc02}) with this ansatz we find that three
non-trivial solutions exist for the choice
\begin{eqnarray}
\omega_i=-\frac{3\alpha^2}{16},\;\;
\delta_i=\gamma_i=\frac{\alpha \beta}{4}\; \mbox{and}\;
\lambda_i=\frac{\alpha \beta}{2},\;i=1,2.
\label{cdvp02}
\end{eqnarray}
The corresponding forms of $R$ and $K$ turn out to be
\begin{eqnarray}
&&(R_1,K_1) = (e^{\frac{3}{4}\alpha t},\;e^{\frac{3}{4}\alpha t}),\quad
(R_2,K_2) = (-\frac{1}{(\dot{y}+\frac{\alpha}{4}y)},\;
\frac{(\dot{x}+\frac{\alpha}{4}x)}{(\dot{y}+\frac{\alpha}{4}y)^2}) \nonumber\\
&&(R_3,K_3) =(\frac{ye^{\frac{1}{4}\alpha t}}{(\dot{y}+\frac{\alpha}{4}y)},\;
-\frac{y(\dot{x}+\frac{\alpha}{4}x)}{(\dot{y}+\frac{\alpha}{4}y)^2}
e^{\frac{1}{4}\alpha t}).
\label{cdvp03}
\end{eqnarray}
Now substituting the form of $R_i$'s and $K_i$'s, $i=1,2,3,$  into
the equations (\ref{eq25}) and (\ref{eq26}), we get the
corresponding $S_i$'s and $U_i$'s, $i=1,2,3$. As a consequence now
we have three sets of independent  solutions for the equations
(\ref{eq23})-(\ref{eq26}) which are all found to be compatible with
(\ref{eq31})-(\ref{eq32}). Substituting the forms
$(S_i,U_i,R_i,K_i)$'s, $i=1,2,3,$ separately into equation
(\ref{cso09}) and evaluating the integrals we get
\begin{align}
I_1& =(\dot{x}+\dot{y}+\frac{\alpha}{4}(x+y)+\frac{\beta}{3}(x+y)^3)
e^{\frac{3}{4}\alpha t},\nonumber\\
I_2&=\frac{(\dot{x}+\frac{\alpha}{4}x)}{(\dot{y}+\frac{\alpha}{4}y)},\quad
I_3=\bigg(\frac{(x\dot{y}-y\dot{x})}{(\dot{y}+\frac{\alpha}{4}y)}\bigg)
e^{\frac{1}{4}\alpha t}.\label{cdvp04}
\end{align}

To illustrate the theory given in the first part of this section we 
consider the first two integrals $I_1$ and $I_2$ given by
equation~(\ref{cdvp04}), and rewrite them in the form (\ref{met13a})
as
\begin{eqnarray}
I_1& =&(\dot{x}+\dot{y}+\frac{\alpha}{4}(x+y))e^{\frac{3}{4}\alpha t}
+\frac{\beta}{3} ((x+y)e^{\frac{1}{4}\alpha t})^3,\nonumber\\
I_2& =&\frac{(\dot{x}+\frac{\alpha}{4}x)e^{\frac{1}{4}\alpha t}}
{(\dot{y}+\frac{\alpha}{4}y)e^{\frac{1}{4}\alpha t}}.
\label{cdvp07}
\end{eqnarray}
Now splitting the first term in $I_1$ and $I_2$ further in the form
(\ref{met13b}) as
\begin{eqnarray}
I_1&=&e^{\frac{1}{2}\alpha t}\bigg(\frac {d}{dt}[(x+y)e^{\frac{1}{4}\alpha t}]\bigg)
+\frac{\beta}{3}(x+y)^3e^{\frac{3}{4}\alpha t},\nonumber\\
I_2& =&\frac{(\dot{x}+\frac{\alpha}{4}x)e^{\frac{1}{4}\alpha t}}
{(\dot{y}+\frac{\alpha}{4}y)e^{\frac{1}{4}\alpha t}}=\frac{e^{-\frac{1}{4}\alpha t}}{(\dot{y}+\frac{\alpha}{4}y)}
\bigg(\frac {d}{dt}(xe^{\frac{1}{4}\alpha t})\bigg),
 \label{cdvp08}
\end{eqnarray}
we identify the new dependent and independent variables from
(\ref{cdvp08}) by using the relations (\ref{met13c}):
\begin{eqnarray}
w_1 = (x+y)e^{\frac{1}{4}\alpha t}, \quad
z_1 = -\frac {2}{\alpha }e^{-\frac{1}{2}\alpha t},\quad
w_2 = xe^{\frac{1}{4}\alpha t}, \quad z_2 = ye^{\frac{1}{4}\alpha t}.
\label{cdvp09}
\end{eqnarray}
One can easily check that in the new variables
equations~(\ref{cdvp04}) become uncoupled as
\begin{eqnarray}
\frac{d^2w_1}{d^2z_1}+\beta w_1^2\frac{dw_1}{dz_1}=0, \quad
\frac{d^2w_2}{d^2z_2}=0.\label{cdvp11a}
\end{eqnarray}

The integrals of (\ref{cdvp11a}) can be easily obtained. These are nothing but the integrals $I_1$ and $I_2$ given in (\ref{cdvp07}) but in terms of the new 
variables (\ref{cdvp09}) they read as
\begin{eqnarray}
I_1 =\frac{dw_1}{dz_1}+\frac{\beta}{3} w_1^3,\qquad
I_2=\frac{dw_2}{dz_2}.\label{cdvp11}
\end{eqnarray}
Solving equations (\ref{cdvp11}),
we obtain (Gradshteyn \& Ryzhik 1980)
\begin{eqnarray}
&&z_1-z_0=\frac {a}{3I_1}\left[\frac{1}{2}
\log\left(\frac{(w_1+a)^2}{w_1^2-aw_1+a^2}\right)
+\sqrt{3}\,\tan^{-1}\left(\frac{w_1\sqrt{3}}{2a-w_1}\right)\right],
\nonumber\\
&& w_2=I_2z_2+I_3,
\label{cdvp14}
\end{eqnarray}
where $a=\sqrt[3]{-\frac {3I_1}{\beta}}$, $I_3$ and $z_0$ are the third and
fourth integration constants. Rewriting $w_i$ and $z_i$, $i=1,2,$ in
terms of the old variables, one can get the explicit solution for
equation (\ref {cdvp01}) for the parametric choice
(\ref{cdvp02}) in the form
\begin{eqnarray}
&&-\frac {2}{\alpha }e^{-\frac{1}{2}\alpha t}-z_0=\frac {a}{3I_1}\bigg[\frac{1}{2}
\log\bigg(\frac{((x+y)e^{\frac{1}{4}\alpha t}+a)^2}{(x+y)^2e^{\frac{1}{2}\alpha t}-a(x+y)e^{\frac{1}{4}\alpha t}+a^2}\bigg)\nonumber\\
&&\qquad\qquad\qquad
+\sqrt{3}\,\tan^{-1}\left(\frac{(x+y)e^{\frac{1}{4}\alpha t}\sqrt{3}}{2a-(x+y)e^{\frac{1}{4}\alpha t}}\right)\bigg],
\nonumber\\
&& x=I_2y+I_3e^{-\frac{1}{4}\alpha t},
\label{cdvp14aa}
\end{eqnarray}
from which one can obtain implicit relations between $x$ and $t$ and
$y$ and $t$, corresponding to the general solution.

\section{Conclusion}
In this paper, as a first part of our investigations on the complete
integrability and linearization of two coupled second order ODEs, we
have focused our attention only on the integrability aspects.  In
particular,  we have introduced a general method of finding integrable
parameters, integrating factors, integrals of motion and general
solution associated with a set of two coupled second order ODEs
through the extended PS procedure. The procedure can also be 
extended straightforwardly to analyse any number of coupled 
second order ODEs. The proposed method is simple,
straightforward and very useful to solve a class of coupled second
order ODEs.  We have illustrated the theory with potentially
important examples.  We have also introduced a novel idea of
transforming coupled second order nonlinear ODEs into uncoupled
second order ODEs.  We have deduced the transformations from the
first two integrals themselves.

It is also of interest to study the problem of linearization of 
coupled nonlinear ODEs by transforming
them into linear ODEs as in the case of single second order 
nonlinear ODEs. However, it is a more difficult and
challenging problem than quadratures.  The primary reasons are (1)
First of all it is not known, in general, whether the given equation
is linearizable or not since in the case of two degrees of freedom
systems there can exist many types of linearizing transformations 
and it is not obvious which one will be successful.  (2)
What are the possible transformations that could exist in the case
of two coupled second order ODEs is also not known in the
literature. (3) Finally, there is no simple procedure that gives us
the required transformation in a straightforward way.  We would like
to address all these questions in the follow up paper (Paper V).

The work of MS forms part of a research project sponsored by National 
Board for Higher Mathematics, Government of India.  The work of ML 
forms part of a Department of Science and Technology, Government 
of India sponsored research project and is supported by a DST 
Ramanna Fellowship.

\end{document}